\documentclass[%
 reprint,
superscriptaddress,
 amsmath,amssymb,
 aps,
pra,
]{revtex4-2}

\usepackage{graphicx}
\usepackage{bm}
\usepackage{amsmath}
\usepackage{hyperref}
\usepackage{esint}
\usepackage{comment,xcolor}
\begin{document}
\title{Time-to-space ghost imaging with classical light}%

\author{Nikita Solonovich} \email{nikita.solonovich@uef.fi}
\affiliation{Department of Physics and Mathematics, University of Eastern Finland, Yliopistokatu 7, Joensuu, 80101, Finland}
\author{Chaoliang Ding} \email{clding@email.tjut.edu.cn}
\affiliation{School of Science, Tianjin University of Technology, Tianjin 300384, China}
\author{Polina P. Kuzhir}
\affiliation{Department of Physics and Mathematics, University of Eastern Finland, Yliopistokatu 7, Joensuu, 80101, Finland}
\author{Tero Set\"al\"a}
\affiliation{Department of Physics and Mathematics, University of Eastern Finland, Yliopistokatu 7, Joensuu, 80101, Finland}
\author{Ari T. Friberg}
\affiliation{Department of Physics and Mathematics, University of Eastern Finland, Yliopistokatu 7, Joensuu, 80101, Finland}
\author{Dmitri B. Horoshko}
\affiliation{
Institute for Quantum Optics, Ulm University, Albert-Einstein-Allee 11, 89081 Ulm, Germany}

\date{\today}

\begin{abstract}
Ghost imaging uses two light beams correlated in the transverse position, time, or frequency to create an image of a spatial, temporal, or spectral object. We propose a scheme of time-to-space ghost imaging for creating a spatial image of a temporal object, enabled by two spatio-temporally correlated light beams. Assuming a spatio-temporal Gaussian Schell model for the description of the source, we obtain analytical expressions for the point-spread function of the system and its temporal resolution. We show how the required source of partially coherent light can be realized by a combination of a diffraction grating and a spatial light modulator. As follows from our analysis, the temporal resolution of a time-to-space imaging system is determined by the duration of the laser pulses used and the transverse coherence length imposed by the spatial light modulator, does not depend on the resolution time of the photodetectors, and can reach the sub-picosecond range. 
\end{abstract}
\maketitle

\section{Introduction}
Ghost imaging is a technique for creating an image of an object using two beams of light correlated in some degrees of freedom. In spatial ghost imaging \cite{Belinskii94}, the image of an object is formed by detecting two correlated optical beams: the test beam passing through the sample is detected by a single-pixel (``bucket'') detector, while the reference beam is detected by a camera with a high spatial resolution. Neither of the two detection records is sufficient for building an image of the sample, which appears only in the correlation function of the two records and is based on strong spatial ($x-x$) correlations between the test and reference beams. Spatial ghost imaging was demonstrated with quantum-entangled photons \cite{Pittman95} and with classically correlated light \cite{Bennink02,Ferri05}, each approach having its advantages: quantum ghost imaging provides better contrast and the possibility of two-color imaging \cite{Chan09,Aspden15}, while classical ghost imaging allows for asymmetry between the two beams leading to low-dose imaging that has important applications in biological and X-ray imaging \cite{Yu16,Pelliccia16}. Ghost imaging attracts much attention due to its inherent insensitivity to the distortion that may occur between the object and the bucket detector \cite{Padgett17,Gilaberte19,Defienne24}, allowing one to form high-resolution images in a strongly scattering medium, e.g. in optical coherence tomography \cite{Amiot19,Huyan22} or none-line-of-sight imaging \cite{Altmann18}.  

In temporal ghost imaging \cite{Shirai10} the image of a temporal object, whose transmittance changes with time, is formed in a similar manner by detecting two temporally correlated optical beams: the test beam passing through the sample is detected by a single-temporal-pixel detector, while the reference beam is detected by a fast detector with high temporal resolution.  Again, the image appears only in the correlation function of the two recorded data sets and relies on strong temporal ($t-t$) correlations between the test and reference beams. This technique was also demonstrated with photon pairs \cite{Denis17}, and classically correlated light \cite{Ryczkowski16,Wu19,Ooka23} and stimulated interest in new potential applications of temporal ghost imaging in the dynamical characterization of free-electron lasers \cite{Ratner19}, secure optical communications \cite{Pan17,Jiang17,He25}, quantum device characterization \cite{Wu20b} and  ultrahigh-frequency signal transmission \cite{Wang22}. The temporal resolution of a temporal imaging system is determined by the response time of the fast detector, and its best reported value is 55 ps \cite{Ryczkowski16}. This value is already at the limit of the temporal resolution of the photodetectors and its improvement is possible by temporal magnification in the reference arm \cite{Ryczkowski17} and by decreasing the correlation time of the beams by employing, e.g., a fiber laser as a source \cite{Wu20a}.

Various similar techniques have also been proposed. Ghost spectroscopy relies on the frequency ($\omega-\omega$) correlations of two beams for the measurement of the object absorption spectrum, as was demonstrated with entangled photons \cite{Scarcelli03} and classically correlated light \cite{Janassek18}. Ghost imaging spectroscopy uses the simultaneous correlation of entangled photons in transverse position and frequency ($\{x,\omega\}-\{x,\omega\}$) to obtain spatially resolved spectra of the object \cite{Chiuri23}. Spatiotemporal ghost imaging has been proposed for simultaneous spatial and temporal imaging on the basis of spatial and temporal ($\{x,t\}-\{x,t\}$) correlations of a classical beam \cite{Abbas20,Abbas20b}. However, the two correlated degrees of freedom should not necessarily be the same. The recently proposed technique of time-to-space ghost imaging \cite{Horoshko23} is based on strong temporal-spatial ($t-x$) correlations between the test and reference beams created in type-I parametric downconversion \cite{Gatti09,Horoshko12,Gatti12,Perina15,LaVolpe20,LaVolpe21,Roux21}. This technique uses spatial measurement of the transverse intensity distribution of the reference beam to create a spatial image of a temporal object probed by the test beam. It avoids the limitation of the detector speed, and its temporal resolution can reach hundreds of femtoseconds in a realistic example.

In this paper, we show how the time-to-space ghost imaging technique of Ref.~\cite{Horoshko23} can be realized with classical spatio-temporally correlated partially coherent beams of light, which we describe by a spatio-temporal Gaussian Schell model (STGSM) including the correlations between the spatial and temporal degrees of freedom. As we show, this model provides an analytical solution for the point-spread function of the imaging system and therefore a simple expression for its temporal resolution. A simple method for generation of the required spatio-temporal correlations is proposed by passing a short pulse of coherent laser light through a combination of a diffraction grating and a spatial light modulator. As follows from our analysis, the temporal resolution of a time-to-space imaging system is generally limited by the duration of the light pulses used. When the pulse duration drops below some limit value, the temporal resolution is determined by the transverse coherence length imposed by the spatial light modulator. In any case, it is not limited by the detector resolution time, in contrast to traditional temporal ghost imaging \cite{Shirai10,Denis17,Ryczkowski16,Wu19}.

The paper is structured as follows. In Sec. II, we introduce a source of partially coherent light, obeying the STGSM that includes possible correlations between the spatial and temporal degrees of freedom. In Sec. III, we show how such a source can be applied to building a spatial image of a temporal object. A method for creating a STGSM source is presented in Sec. IV, where the limitations on the temporal resolution are discussed. Section V summarizes the results and concludes the paper.

\section{Spatio-temporal Gaussian Schell model \label{sec:intro}}

Consider a source of partially coherent light, where the positive frequency part of the electric field $E_0(\boldsymbol{\xi})$ is a function of the transverse position $x$ and the time $t$, which we write as a column vector $\boldsymbol{\xi} = (x,t)^T$ with the superscript $T$ standing for transposition. The mutual coherence function \cite{Mandel&Wolf} of the light beam is defined as the cross-correlation function of the electric field at two spatial points $\Gamma_0(\boldsymbol{\xi}_1,\boldsymbol{\xi}_2) = \langle E^*_0(\boldsymbol{\xi}_1)E_0(\boldsymbol{\xi}_2)\rangle$. Extending the general formalism of Schell model \cite{Mandel&Wolf} to the spatio-temporal case, we represent it as
\begin{equation}\label{Gamma0}
\Gamma_0(\boldsymbol{\xi}_1,\boldsymbol{\xi}_2) = \left[\langle I(\boldsymbol{\xi}_1)\rangle\langle I(\boldsymbol{\xi}_2)\rangle\right]^\frac12 \gamma(\boldsymbol{\xi}_1-\boldsymbol{\xi}_2),
\end{equation}
where $I(\boldsymbol{\xi}) = E_{0}^*(\boldsymbol{\xi})E_{0}(\boldsymbol{\xi})$ is the field intensity and $\gamma(\boldsymbol{\xi}_1-\boldsymbol{\xi}_2)$ is the complex degree of coherence of light at the points $\boldsymbol{\xi}_1$ and $\boldsymbol{\xi}_2$. In a spatio-temporal Gaussian Schell model, both the mean intensity and the complex degree of coherence are assumed to be Gaussian functions of their arguments. Thus, we express the mean intensity as the most general Gaussian function of two arguments in a form
\begin{equation}\label{I}
   \langle I(\boldsymbol{\xi})\rangle = I_0\exp\left(-\frac12\boldsymbol{\xi}^\dagger\boldsymbol{\Sigma}_I^{-1}\boldsymbol{\xi}\right),
\end{equation}
where $I_0$ is a positive constant, $\dag$ stands for Hermite conjugation, and
\begin{equation}\label{SigmaI}
   \boldsymbol{\Sigma}_I = \left[\begin{array}{cc} w^2 & \rho_I wT \\ \rho_I wT & T^2\end{array}\right]
\end{equation}
is the mean intensity correlation matrix that includes the spatial and temporal standard deviations of the pulsed light beam $w$ and $T$ respectively, and the dimensionless correlation coefficient $\rho_I$. The integrability of $I(\boldsymbol{\xi})$ requires the positivity of matrix $\boldsymbol{\Sigma}_I$, which, in its turn, requires that $|\rho_I|<1$. The reciprocal of the mean intensity correlation matrix is given by
\begin{equation}\label{SigmaIres}
   \boldsymbol{\Sigma}_I^{-1} = \frac1{1-\rho_I^2}\left[\begin{array}{cc} 1/w^2 & -\rho_I/wT \\ -\rho_I/wT & 1/T^2\end{array}\right].
\end{equation}

Analogously, we write the modulus of the complex degree of coherence as the most general Gaussian function of two arguments in a form
\begin{equation}\label{gammamod}
   |\gamma(\boldsymbol{\xi})|= \exp\left(-\frac12\boldsymbol{\xi}^\dagger\boldsymbol{\Sigma}_\gamma^{-1}\boldsymbol{\xi}\right),
\end{equation}
where 
\begin{equation}\label{SigmaG}
   \boldsymbol{\Sigma}_\gamma = \left[\begin{array}{cc} \sigma^2 & \rho_\gamma \sigma\tau \\ \rho_\gamma \sigma\tau & \tau^2\end{array}\right]
\end{equation}
is the correlation matrix that includes the spatial and temporal standard deviations of the field correlations in space and time $\sigma$ and $\tau$, respectively, and the dimensionless correlation coefficient $\rho_\gamma$. Similarly, the integrability of the complex degree of coherence requires $|\rho_\gamma|<1$ and the reciprocal of the correlation matrix is
\begin{equation}\label{SigmaGres}
   \boldsymbol{\Sigma}_\gamma^{-1} =\frac1{1-\rho_\gamma^2} \left[\begin{array}{cc} 1 / \sigma^2 & -\rho_\gamma / \sigma\tau \\ -\rho_\gamma / \sigma\tau & 1 / \tau^2\end{array}\right].
\end{equation}
The phase of a Gaussian complex degree of coherence can, in principle, be a second-order polynomial of $x$ and $t$. However, in this work, for the sake of simplicity, we assume it to be a linear function of time only and write
\begin{equation}\label{gamma}
   \gamma(\boldsymbol{\xi})=  |\gamma(\boldsymbol{\xi})|e^{i\omega t},
\end{equation}
where $\omega$ has the meaning of the central circular frequency of the beam. A possible way of creating such a source is presented in Sec.~\ref{sec:Generation}.

\section{Ghost imaging}

\subsection{General case}

\begin{figure*}
    \centering
    \includegraphics[width=\linewidth]{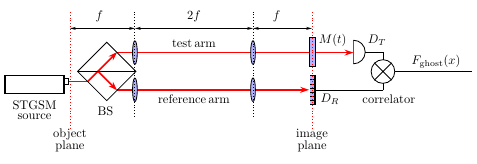}
    \caption{Time-to-space ghost imaging scheme. Spatio-temporally correlated light pulse is split by the beam splitter (BS). In the test arm, after passing through a $4f$ imaging system, the light pulse goes through the temporal object, whose intensity trasmittance $M(t)$ varies with time, and is registered by the bucket detector $D_T$. In the reference arm, the light pulse is imaged on the camera $D_R$ by another $4f$ imaging system. Ghost image $F_\text{ghost}(x)$ of the temporal object appears when the outputs from detectors are correlated. Note that imaging in each arm can include beam expansion or compression (not shown), but both the temporal object and the camera are placed in the image planes of their corresponding $4f$ spatial imaging systems.}
    \label{fig:setup}
\end{figure*}

Following the approaches to spatial \cite{Gatti04} and temporal \cite{Shirai10} ghost imaging with classical light, we consider splitting the source field, described in the preceding section, into two beams on a symmetric beam splitter creating the test and reference beams, see Fig.~\ref{fig:setup}. Each of the two beams travels through its proper imaging system, described by its impulse response function $h_T(\boldsymbol{\xi},\boldsymbol{\xi}')$ or $h_R(\boldsymbol{\xi},\boldsymbol{\xi}')$, for the test and reference beams, respectively. The test arm includes a temporal object, a spatially homogeneous body with a time-varying field transmittance $m(t)$. We assume that the object is amplitude-only, i.e. $m(t)$ is real-valued. After passing through the object, the test beam is detected by $D_T$, which is a bucket detector that collects all the light in the detection plane and integrates its intensity over the response time, assumed to be much longer than the pulse duration $T$. In the reference arm, the camera (detector $D_R$), placed in the object plane, spatially resolves the light intensity in the transverse plane, also integrating it over the response time much longer than $T$. Only one dimension of the camera is used, so a pixel array is sufficient. We assume that the number of illuminated pixels is high enough to describe the pixel position by a continuous variable $x$. The transformation from the field of the source $E_0(\boldsymbol{\xi}')$ to the field in the object plane in the corresponding arm, $E_i(\boldsymbol{\xi}')$, where $i\in\{T,R\}$, reads
\begin{equation}
E_i(\boldsymbol{\xi}) = \int h_i(\boldsymbol{\xi},\boldsymbol{\xi}')E_0(\boldsymbol\xi') d \boldsymbol\xi',
\end{equation}
where integration over $\boldsymbol\xi'$ implies a double integration in $x'$ and $t'$. Here and below, the integration limits are assumed to extend from $-\infty$ to $+\infty$. Both impulse response functions are assumed to be causal, so that $h_i(\boldsymbol{\xi},\boldsymbol{\xi}')=0$ for $t'>t$.

The electric signals from $D_T$ and $D_R$ are multiplied by the correlator. For the $\ell$'th pulse, the bucket detector measures the number of photons in the test arm 
\begin{equation}\label{N}
    N_\ell = \int I_T^\ell(\boldsymbol{\xi})d\boldsymbol{\xi},
\end{equation}
while the camera measures the photon number density in the reference arm
\begin{equation}\label{nx}
    n_\ell(x) = \int I_R^\ell(\boldsymbol{\xi})dt,
\end{equation}
where $I_i(\boldsymbol{\xi})=E_i^*(\boldsymbol{\xi})E_i(\boldsymbol{\xi})$ is the intensity in the respective arm and the superscript $\ell$ denotes the $\ell$th sample of this random process. The ghost image is formed by computing for the $K$ detected pulses the correlation function between the electric signals of the two arms upon subtracting the background of the camera \cite{Katz09}
\begin{equation}
 F_\text{ghost}(x) =  \frac1K\sum\limits_{\ell=1}^K\left(n_\ell(x)-\frac1K\sum\limits_{\ell=1}^K n_\ell(x)\right)N_\ell.  
\end{equation}
Replacing the averaging over subsequent samples by the ensemble averaging in the limit of large $K$ and substituting Eqs. (\ref{N}) and (\ref{nx}), we rewrite this expression as
\begin{equation}\label{image}
F_\mathrm{ghost}(x_2) = \int d\boldsymbol{\xi}_1\int dt_2
G(\boldsymbol{\xi}_1,\boldsymbol{\xi}_2),
\end{equation}
where $G(\boldsymbol{\xi}_1,\boldsymbol{\xi}_2) =
\langle I_T(\boldsymbol{\xi}_1)I_R(\boldsymbol{\xi}_2)\rangle - \langle I_T(\boldsymbol{\xi}_1)\rangle\langle I_R(\boldsymbol{\xi}_2)\rangle$ is the correlation function of the intensity fluctuations in the two arms.  

To calculate $G(\boldsymbol{\xi}_1,\boldsymbol{\xi}_2)$, we need to determine the higher-order correlators of the field in addition to the second-order correlator of our STGSM model, Eq. (\ref{Gamma0}). We assume that the light emitted by the source possesses Gaussian statistics, as is the case for pseudothermal sources typically used in spatial and temporal imaging schemes \cite{Bennink02,Ferri05,Ryczkowski16,Starovoitov24}. It means that the characteristic functional of the source field has a Gaussian form  specified by the phase-insensitive correlator, Eq. (\ref{Gamma0}), and the phase-sensitive correlator $\langle E_0(\boldsymbol{\xi}_1)E_0(\boldsymbol{\xi}_2)\rangle$ \cite{Erkmen08}.
The fourth-order field correlators entering $G(\boldsymbol{\xi}_1,\boldsymbol{\xi}_2)$ can then be decomposed into a sum over the products of all possible second-order correlators, according to the Gaussian moment factoring theorem \cite{Mandel&Wolf}. Although a classical field with a non-zero phase-sensitive correlator can be created \cite{Erkmen08}, we assume that the phase-sensitive correlator of the source field $\langle E_0(\boldsymbol{\xi}_1)E_0(\boldsymbol{\xi}_2)\rangle$ is identically zero, as typical for pseudo-thermal sources. Thus, we arrive, similarly to Ref. \cite{Gatti04}, at a simple expression $G(\boldsymbol{\xi}_1,\boldsymbol{\xi}_2) = \left|\Gamma_{12}(\boldsymbol{\xi}_1,\boldsymbol{\xi}_2)\right|^2$, where 

\begin{align}\label{Gamma12}
&\Gamma_{12}(\boldsymbol{\xi}_1,\boldsymbol{\xi}_2) 
\\
\nonumber
&= \int d\boldsymbol{\xi}_1'\int d\boldsymbol{\xi}_2'
   h_T^*(\boldsymbol{\xi}_1,\boldsymbol{\xi}_1')  h_R(\boldsymbol{\xi}_2,\boldsymbol{\xi}_2')    \Gamma_0(\boldsymbol{\xi}_1',\boldsymbol{\xi}_2')
\end{align}
is the cross-correlation function of two fields in the image plane. 

In the following, we consider near-field ghost imaging \cite{Gatti04,Ferri05}, realized with 4\emph{f} imaging systems in both arms, which correspond to the following impulse-response functions 
\begin{eqnarray} \label{h1}
h_T(\boldsymbol{\xi}_1,\boldsymbol{\xi}_1') &=& \sqrt{\eta_T}\delta(x_1-x_1')\exp\{-ix_1^2\pi/\lambda f\}  \\\nonumber
&\times& m(t_1')\delta(t_1-t_1'-L/c),\\\label{h2}   h_R(\boldsymbol{\xi}_2,\boldsymbol{\xi}_2') &=& \sqrt{\eta_R} \delta(x_2-x_2')\exp\{-ix_2^2\pi/\lambda f\}\\\nonumber
&\times& \delta(t_2-t_2'-L/c),
\end{eqnarray}
where $L$ is the length of each arm, $c$ is the speed of light in vacuum, $f$ is the focal length of the lenses, $\lambda=2\pi c/\omega$ is the central wavelength, and $\eta_i$ is the total intensity transmittance of the corresponding arm, including the split ratio of the beam splitter and the quantum efficiency of the detector. To simplify the formulas, the spatial inversion introduced by a 4\emph{f} system is compensated by inverting the $x$ axis on the image plane. Here, we disregard the diffractive broadening of the impulse-response function, assuming that it is much smaller than the transverse coherence length $\sigma$. 

Substituting the above impulse-response functions into Eqs. (\ref{image}) and (\ref{Gamma12}) and performing the integrations over the delta functions, we obtain
\begin{equation}\label{image2}
    F_\mathrm{ghost}(x_2) = \int  P(x_2|\tilde{t}_1)M(\tilde{t}_1) d\tilde{t}_1,
\end{equation}
where $M(\tilde{t}_1)=|m(\tilde{t}_1)|^2$ is the intensity transmittance of the object and
\begin{equation}\label{P21}
P(x_2|\tilde{t}_1) = \eta_T\eta_R\int d x_1 \int d\tilde t_2 \left|\Gamma_0(\tilde{\boldsymbol{\xi}}_1,\tilde{\boldsymbol{\xi}}_2)\right|^2    
\end{equation}
is the point spread function (PSF) of the time-to-space imaging system with $\tilde{t}_i=t_i-L/c$ and $\tilde{\boldsymbol{\xi}}_i=(x_i,\tilde t_i)^T$. For the STGSM introduced by Eq. (\ref{Gamma0}), the PSF is 
\begin{eqnarray} \label{psf_Gaussian}
P(x_2|\tilde{t}_1)
&=& \eta_T\eta_R I_0^2 \int e^{-\frac12\mathbf{u}^T \mathbf{R} \mathbf{u}+\mathbf{a}^T\mathbf{u}+b}d \boldsymbol{u} \\\nonumber
&=& \frac{2\pi\eta_T\eta_R I_0^2}{\sqrt{\det\mathbf{R}}} 
e^{\frac12\mathbf{a}^T \mathbf{R}^{-1}\mathbf{a}+b},
\end{eqnarray}
where we have used the multidimensional Gaussian integration formula \cite{Srivastava23a} for $\boldsymbol{u}= \left( x_1, \tilde{t}_2 \right) ^T$, while the matrix $\mathbf{R}$, the column vector $\mathbf{a}$ and the scalar $b$ are defined as follows:
\begin{equation}\label{Lambda}
\mathbf{R} = 
\begin{bmatrix} 
 \frac{1}{w^2(1-\rho_I^2)}+\frac{2}{\sigma^2(1-\rho_\gamma^2)} & \frac{2\rho_\gamma}{\sigma\tau (1-\rho_\gamma^2)} \\
 \frac{2\rho_\gamma}{\sigma\tau(1-\rho_\gamma^2)} & \frac{1}{T^2(1-\rho_I^2)} + \frac{2}{\tau^2(1-\rho_\gamma^2)}
\end{bmatrix},
\end{equation}
\begin{eqnarray}\label{a}
    \mathbf{a} &=& \biggl\{ \frac{\rho_I \tilde{t}_1}{T w (1-\rho_I^2)} +\frac{2 x_2}{\sigma^2 (1-\rho_\gamma^2)}+\frac{2 \rho_\gamma \tilde{t}_1}{\sigma \tau (1-\rho_\gamma^2)},
    \\ \nonumber
    &\,&\frac{\rho_Ix_2}{Tw(1-\rho_I^2)} + \frac{2\tilde{t}_1}{\tau^2(1-\rho_\gamma^2)} + \frac{2 \rho_\gamma x_2}{\sigma \tau (1-\rho_\gamma^2)}
    \biggr\},
\end{eqnarray}
\begin{eqnarray}\label{b}
    b &=& -x_2^2\left[ \frac{1}{2w^2\left(1-\rho_I^2\right)}+\frac{1}{\sigma^2\left(1-\rho_\gamma^2\right)}\right]
    \\ \nonumber
    &-& \tilde{t}_1^2 \left[\frac{1}{2T^2\left(1-\rho_I^2\right)}+\frac{1}{\tau^2\left(1-\rho_\gamma^2\right)}\right]
    \\ \nonumber
    &-& \tilde{t}_1 x_2  \frac{2\rho_\gamma}{\sigma\tau \left(1 - \rho_\gamma^2\right)}.
\end{eqnarray}

It can be easily seen from Eqs.~(\ref{Lambda}), (\ref{a}), and (\ref{b}), that the exponent on the right-hand side of Eq.~(\ref{psf_Gaussian}) is a second-order polynomial in $\tilde{t}_1$ and $x_2$, which can be written as
\begin{equation}\label{poly}
\frac12\mathbf{a}^T \mathbf{R}^{-1}\mathbf{a}+b = 
-C_{tt}\tilde{t}_1^2 - 2C_{tx}\tilde{t}_1x_2 - C_{xx}x_2^2,
\end{equation}
where $C_{ij}$ are some coefficients. The integrability of the PSF requires that the eigenvalues of the coefficient matrix $C_{ij}$ are positive, and consequently the trace and determinant of this matrix are positive, that is $C_{tt}+C_{xx}>0$, and $C_{tt}C_{xx}-C_{tx}^2>0$. When these conditions are satisfied, we can rewrite the PSF in the form
\begin{equation}\label{generalPSF}
   P(x_2|\tilde{t}_1) = P_\mathrm{max}\exp\left[-\frac{(\tilde{t}_1-x_2/v)^2}{2\tau_\text{PSF}^2}-\frac{x_2^2}{2w_\text{PSF}^2}\right],
\end{equation}
where $v = - C_{tt}/C_{tx}$, $\tau_\text{PSF}^2 = 1/2C_{tt}$,
\begin{eqnarray}
w_\text{PSF}^2 &=& \frac{C_{tt}}{2(C_{tt}C_{xx}-C_{tx}^2)},\\
P_\text{max} &=& \frac{2\pi\eta_T\eta_R I_0^2}{\sqrt{\det \mathbf{R}}}.
\end{eqnarray}

To clarify the physical meaning of the parameters introduced, we consider a point-like temporal object centered at $\tilde{t}_1=t_0$, which we model by a delta function, $M(\tilde{t}_1)=\delta(\tilde{t}_1-t_0)$. Physically, such an object can be created by passing a Gaussian sub-picosecond-scale theraherz pulse through an electrooptical light modulator, which will result in the opening of the latter at $t_0$ for a short time much less than the duration of the optical pulses used in the experiment. From Eq.~(\ref{image2}), we see that in this case the ghost image is given by $P(x_2|t_0)$, which means that the intensity distribution on the camera represents a Gaussian spot centered at $vt_0$ and having a standard deviation of approximately $v\tau_\text{PSF}$ in the regime $w_\text{PSF}\gg v\tau_\text{PSF}$.
Therefore, the parameter $v$, having a velocity dimension, shows the sensitivity of the ghost image to the temporal shift of the object. Alternatively, it can be understood as a magnification of the time-to-space imaging system. The spot standard deviation $v\tau_\text{PSF}$ can be understood as the output resolution of the system. Dividing the output resolution $v\tau_\text{PSF}$ by the magnification $v$, we obtain the input temporal resolution $\tau_\text{PSF}$. The peak intensity of the spot is $P_\text{max}\exp\left(-v^2t_0^2/2w_\text{PSF}^2\right)$, which means that $\Delta t=w_\text{PSF}/v$ determines the field of view of the imaging system. The general expressions for these parameters are rather cumbersome and can be found in Appendix~\ref{sec:app0}. In the next section, we consider a particular case of full temporal coherence, which provides a rather simple analytical description and a feasible experimental implementation.

\subsection{Case of full temporal coherence}

\begin{figure}[ht!]
\centering
\includegraphics[width=\linewidth]{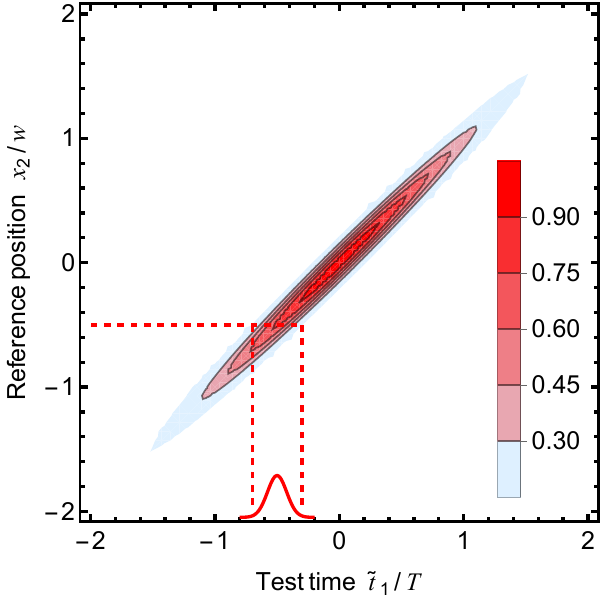}
\caption{Normalized PSF $P(x_2|\tilde{t}_1)/P_\text{max}$ of the time-to-space ghost imaging system with $q^2/2+\varepsilon^2=0.01$. Each position $x_2$ corresponds to a temporal interval which is characterized by Gaussian distribution over time with the standard deviation $\tau_\text{PSF}$ constituting the temporal resolution.
\label{fig:PSF}}
\end{figure}

In this section, we consider a relatively simple particular case of full temporal coherence of the source field corresponding to $\tau=\infty$. In this case, $\sigma$ and $\rho_\gamma$ enter the expressions for $\mathbf{R}$, $\bf{a}$, and $b$ only in a combination $\sigma\sqrt{1-\rho_\gamma^2}$, i.e., the role of the correlation between spatial and temporal coherence is only in reducing the coherence length. For this reason, in this section we assume $\rho_\gamma=0$. With these simplifications, we rewrite Eqs. (\ref{Lambda}), (\ref{a}), and (\ref{b}) as follows:
\begin{eqnarray}\label{LambdaTinf}
\mathbf{R} &=& 
\begin{bmatrix} 
 \frac{1}{w^2\varepsilon^2}+\frac{2}{\sigma^2} & 0 \\
 0 & \frac{1}{T^2\varepsilon^2}
\end{bmatrix},
\\ 
\label{aTinf}
    \mathbf{a} &=& \left( \frac{\rho_I \tilde{t}_1}{T w \varepsilon^2} +\frac{2 x_2}{\sigma^2},\,\, \frac{\rho_Ix_2}{T w \varepsilon^2}
    \right),
\\
\label{bTinf}
    b &=& -x_2^2\left( \frac{1}{2w^2\varepsilon^2}+\frac{1}{\sigma^2}\right)
    - \frac{\tilde{t}_1^2}{2T^2\varepsilon^2},
\end{eqnarray}
where we have introduced a new parameter $\varepsilon=\sqrt{1-\rho_I^2}$, which can be understood as the ``degree of decorrelation'' of time and lateral position in the mean intensity of the source field.
Substituting Eqs.~(\ref{LambdaTinf})-(\ref{bTinf}) into Eq.~(\ref{psf_Gaussian}), we obtain the PSF given by Eq.~(\ref{generalPSF}) with the following parameters (see Appendix \ref{sec:app1}):
\begin{eqnarray}\label{v}
v&=& \frac{\sigma^2+2w^2}{2Tw\rho_I}\rightarrow \frac{w}{T},\\\label{tauPSF}
\tau_\text{PSF}&=&T\sqrt{\frac{\sigma^2+2w^2\varepsilon^2}{\sigma^2 + 2w^2}}\rightarrow T\sqrt{\frac{q^2}2 + \varepsilon^2},\\\label{w}
w_\text{PSF}&=& w\sqrt{\frac{\sigma^2+2w^2}{\sigma^2+4w^2}}\rightarrow \frac w{\sqrt2},\\\label{Pmax}
P_\text{max} &=&  \frac{2\pi\eta_T\eta_R I_0^2 T\sigma\varepsilon^2}{\sqrt{q^2+2\varepsilon^2}},
\end{eqnarray}
where we have introduced the degree of global (spatial) coherence $q=\sigma/w$ \cite{Mandel&Wolf} and the approximate formulas correspond to the limit of low spatial coherence and high spatiotemporal correlation, $q,\varepsilon\ll1$. Dividing the field of view $\Delta t=w_\text{PSF}/v$ by the resolution time, we obtain the number of resolved features
\begin{equation}\label{Nf}
N_f = \frac{\Delta t}{\tau_\text{PSF}} \to \frac1{\sqrt{q^2+2\varepsilon^2}}.
\end{equation}
An example of PSF in this limit is shown in Fig.~\ref{fig:PSF}.

Equation~(\ref{tauPSF}) shows that the resolution time can be made much shorter than the pulse duration of the source field $T$ by reducing the parameters $q$ and $\varepsilon$ much below 1, which shows the power of the time-to-space ghost imaging system for high-temporal-resolution imaging. However, $T$ is the duration of the light pulse with engineered spatiotemporal correlations and partial spatial coherence and it can be much longer than the duration of the initial laser pulse $T_0$ used for such engineering. It would be interesting to understand the resolution limit with respect to the initial laser pulse.  In the next section, we consider a possible scheme for transforming a Gaussian laser pulse with duration $T_0$ and width $w_0$ into an STGSM pulse described in Sec.~\ref{sec:intro}, which allows us to obtain the ultimate limit for the resolution of the time-to-space ghost imaging technique.

\section{Generation of spatiotemporally correlated partially coherent light \label{sec:Generation}}

For a light pulse with the mean intensity given by Eq.~(\ref{I}), the maximal intensity at given time $t$ corresponds to the point where $\partial \boldsymbol{\xi}^\dagger\boldsymbol{\Sigma}_I^{-1}\boldsymbol{\xi}/\partial x=0$, which gives $x=\rho_Iwt/T$. This means that the intensity front of the pulse is tilted with respect to the phase front, which is perpendicular to the propagation direction. Pulse-front tilt has various applications in optics and can be produced by a dispersive optical element such as a prism or a diffraction grating \cite{Toth23}. We consider a laser pulse propagating along the direction $z'$ and having the field distribution at $z'=0$
\begin{equation}
E_L(x',t)=E_L^\text{max}e^{-{x'}^2/4w_0^2 -t^2/4T_0^2-i\omega t}, 
\end{equation}
where $x'$ is the transverse position with respect to the direction $z'$, $w_0$ and $T_0$ are the standard deviations of the intensity distribution in space and time, respectively. Since this field is fully coherent, its mean intensity is simply $\langle I_L(x',t)\rangle=\left|E_L(x',t)\right|^2$. For such a pulse, the spatial and temporal degrees of freedom are  not correlated because the mean intensity factors out into a product of a function of $x'$ and a function of $t$.

\begin{figure}[ht!]
\centering
\includegraphics[width=\linewidth]{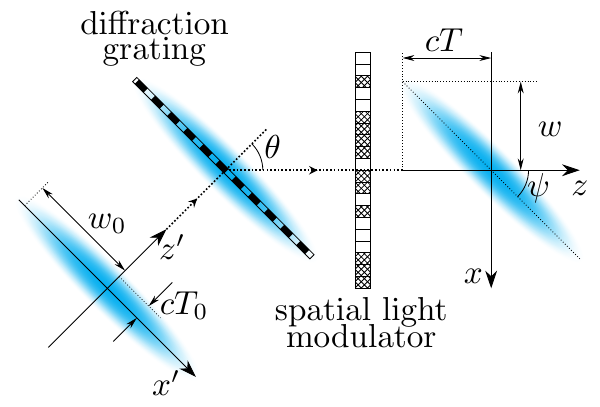}
    \caption{Scheme of an STGSM light source. The diffraction grating produces a tilt between the intensity and wave fronts by angle $\psi=\theta$. Duration of the pulse increases $T_0\to T$ while its width decreases $w_0\to w$ as a result of the wavefront rotation. The  spatiotemporal correlation coefficient is zero before the grating and is $\rho_I\approx1$ in the diffracted light. The spatial light modulator is driven by Gaussian noise and used to reduce the transverse coherence length to the smallest possible value $\sigma$ determined by the pixel size.}
\label{fig:grating}
\end{figure}

This pulse impinges on a diffraction grating and the diffracted beam of order $m_d$ is selected, as shown in Fig.~\ref{fig:grating}. We consider a transmission grating, but a reflection grating will produce the same result. The full transformation of a Gaussian pulse focused at a distance $d$ from the grating and observed at a distance $z$ from it can be found in Ref.~\cite{Martinez86}. Since we consider the near-field ghost imaging, we disregard the diffraction at the beam propagation and adopt the expression for the diffracted field at $d=z=0$, which gives 
\begin{equation}\label{ED}
E_D(x,t)=E_D^\text{max}e^{-\alpha^2x^2/4w_0^2 -(t-x\tan\psi/c)^2/4T_0^2-i\omega t},    
\end{equation}
where $x$ is the transverse position with respect to the direction of diffracted beam propagation $z$ and the parameter $\alpha$ and the tilt angle $\psi$ are defined as follows, 
\begin{eqnarray}\label{alpha}
\alpha &=& -\frac{\cos\varphi}{\cos\theta},\\\label{beta}
\tan\psi &=& \frac{m_d\lambda}{ \Lambda\cos\theta},
\end{eqnarray}
with $\varphi$ the angle of incidence, $\theta$ the angle of diffraction, and $\Lambda$ the period of the grating. Note that our definition of the tilt angle includes a minus sign with respect to that of Martinez~\cite{Martinez86} so that a positive tilt angle corresponds to a positive diffraction angle.

Equations~(\ref{ED}), (\ref{alpha}), and (\ref{beta}), supplemented by the grating equation \cite{Goodman-Book} $\sin\theta = \sin\varphi + m_d\lambda/\Lambda$, are sufficient to find the matrix $\boldsymbol{\Sigma}_I$ of diffracted light. However, to avoid cumbersome expressions, we consider the case of normal incidence $\varphi=0$, as shown in Fig.~\ref{fig:grating}. In this case, the tilt angle is equal to the angle of diffraction, $\psi=\theta$, and the mean intensity of diffracted light, $\langle I_D(x,t)\rangle=\left|E_D(x,t)\right|^2$, can be written in the form of Eq.~(\ref{I}) with
\begin{equation}\label{SigmaIresbis}
   \boldsymbol{\Sigma}_I^{-1} = \frac1{c^2T_0^2}\left[\begin{array}{cc} \delta^2 \cos^{-2}\theta + \tan^2\theta & -c\tan\theta \\ -c\tan\theta & c^2\end{array}\right],
\end{equation}
where $\delta=cT_0/w_0$. Comparing this expression with Eq.~(\ref{SigmaIres}), we find a system of three equations
\begin{eqnarray}
\frac{1}{\varepsilon^2 w^2}&=&\frac{\delta^2\cos^{-2}\theta+\tan^2\theta}{c^2 T_0^2},
    \\
\frac{\rho_I}{\varepsilon^2 wT}&=&\frac{\tan\theta}{cT_0^2},
    \\
\frac{1}{\varepsilon^2 T^2}&=&\frac1{T_0^2}.
\end{eqnarray}
Solving this system, we obtain
\begin{eqnarray}\label{rhoI_grating}
\rho_I &=& \frac{\sin\theta}{\sqrt{\delta^2+\sin^2\theta}},
\\ \label{Pulse_duration_after_grating}
T&=&\sqrt{T_0^2+w_0^2\sin^2\theta/c^2},
\\\label{wgrating}
w&=&w_0\cos\theta.
\end{eqnarray}
We see that the modulus of the intensity correlation coefficient $\rho_I$ tends to its maximum value of 1 for a ``thin'' pulse whose length $cT_0$ is much less than its width $w_0$. 

The complex degree of coherence with its modulus given by Eq.~(\ref{gammamod}) can be created by passing the tilted pulse through a rotating ground glass disk. In this technique, one typically creates partially coherent light in the far field of the disk by means of a lens with focal length $f_L$, in which case the coherence length is $\sigma=\lambda f_L/\pi w\sqrt{2}$ \cite{DeSantis79} and can be varied by changing $f_L$ or $w$ \cite{Starovoitov23}. However, the same effect can be reached in the near field of the disk \cite{Wang06}, in which case the coherence length is fixed to the average size of the disk inhomogeneities. Alternatively, partial coherence can be imposed on a coherent beam by means of a spatial light modulator \cite{Hyde15,Cai17}, as shown in Fig.~\ref{fig:grating}.

Now we can find the temporal resolution of the time-to-space ghost imaging system in terms of the initial laser pulse. Substituting Eqs.~(\ref{rhoI_grating}) -- (\ref{wgrating}) into Eq.~(\ref{tauPSF}), we obtain 

\begin{equation}\label{resolution}
\tau_\text{PSF} = T_0 \sqrt{1+\frac{\sigma^2}{2w_0^2\cos^2\theta}+\frac{\sigma^2}{2c^2T_0^2}\tan^2\theta}.
\end{equation}

From Eq.~(\ref{resolution}), we can see that the duration of the laser pulse $T_0$ is the lower limit of temporal resolution in a time-to-space ghost imaging system. This limit is reached when two conditions are satisfied:  $\sigma\ll w_0\cos\theta$ and $\sigma\ll cT_0/\tan\theta$. In an experiment, the transverse coherence length $\sigma$ is determined by the average size of the inhomogeneities in the ground glass disk or by the pixel size of a spatial light modulator and cannot typically be modified. For a not too high order of diffraction, both $\cos\theta$ and $\tan\theta$ are of the order of unity. Therefore, the two conditions above can be rewritten as $w_0\gg \sigma$ and $T_0\gg\sigma/c$ and understood as limitations on the spatial and temporal size of the laser pulse focused on the diffraction grating. The first condition is rather easy to reach by a not too tight focusing. However, the second condition limits $T_0$ and therefore the temporal resolution. When the first condition is satisfied, we can consider  $\tau_\text{PSF}$ as a function of $T_0$. This function reaches its minimum value 
\begin{equation}\label{taumin}
\tau_\text{PSF}^\text{min} = \frac{\sigma\tan\theta}{\sqrt{2}c}
\end{equation}
for $T_0\to0$. Taking a realistic value of $\sigma=35$ $\mu$m \cite{Ferri05} and $\tan\theta\approx1$, we obtain $\tau_\text{PSF}^\text{min} = 83$ fs. Substituting Eqs.~(\ref{rhoI_grating}) -- (\ref{wgrating}) into Eq.~(\ref{Nf}), we obtain the number of resolved features
\begin{equation}\label{Nfbis}
N_f = \frac1\sigma\sqrt{w_0^2\cos^2\theta+2\varepsilon^2},
\end{equation}
which is approximately equal to the number of illuminated SLM pixels $w_0\cos\theta/\sigma$. 

To reach a low resolution time predicted by Eq.~(\ref{taumin}), one needs to satisfy the conditions implied in its derivation. First, in addition to the tilt of the pulse front, the diffracted light experiences angular dispersion, that is, each spectral component diffracts at a different angle, leading to a significant lateral walkoff for an ultrashort pulse that has an ultrabroad spectrum. This effect can be neglected if the distance between the grating and the SLM is much less than the walkoff distance \cite{Martinez86} 
\begin{equation}
z_\text{walkoff}=\frac{4\pi cT_0w_0}\lambda \frac{\cos\theta}{\tan\theta}.    
\end{equation}
Using the values of $\lambda=500$ nm, $\theta=45^{\circ}$, $T_0=50$ fs, $w_0=1$ mm, we obtain $z_\text{walkoff}\approx25$ cm.
If the distance between the grating and the SLM cannot be much less than this distance, another $4f$ imaging system is required to image the grating on the surface of the SLM, providing a further possibility of manipulating the tilt angle and the beam waist \cite{Toth23}. 

Second, the condition of high spatio-temporal correlation, $\varepsilon\ll1$, implies, according to Eq.~(\ref{rhoI_grating}), that
\begin{equation}
T_0 \ll \frac{w_0\sin\theta}{c},   
\end{equation}
which means that the angle of diffraction $\theta$ cannot be too small. Upon satisfying these two conditions and using ultrashort laser pulses (duration 50 fs or less), one can obtain a temporal resolution as low as 83 fs, which is more than two orders of magnitude better than the resolution of traditional temporal ghost imaging \cite{Shirai10,Denis17,Ryczkowski16,Wu19}, where the resolution time is 55 ps or higher.

\section{Conclusion}
In this paper, we have introduced the STGSM for partially coherent pulsed light, which is a generalization of the spatial Gaussian Schell model \cite{Mandel&Wolf} to the case of spatio-temporal correlations. Although the generalization is straightforward, we are unaware of a similar work. In addition, we have described a method for creating an STGSM source in the limit of infinite temporal coherence by passing a short pulse of coherent laser light through a combination of a diffraction grating and an SLM.

Using this model for the source of partially coherent light, we have shown that a time-to-space ghost imaging scheme can be built on its basis, that allows one to create a spatial ghost image of a temporal object. The key advantage of this technique is very short resolution time, on the order of 100 fs, two orders of magnitude below that of the conventional temporal ghost imaging limited by the response time of the photodetectors. 

It should be noted that the spatial ghost imaging technique has often been replaced in the last decade by the computational ghost imaging technique \cite{Altmann18}, where the reference beam is omitted and the test beam is spatially modulated by a digital micromirror device (DMD) creating random spatial patterns for object illumination. A similar technique has also been developed for computational temporal ghost imaging reaching sub-picosecond temporal resolution in the near-infrared \cite{Zhao21} and 80 ps resolution in the mid-infrared \cite{Zhang26} spectral ranges. Since the degree of complexity and price of a DMD \cite{Zhao21} or a pair of nonlinear frequency converters \cite{Zhang26}  are comparable to those of a camera, we believe that our time-to-space ghost imaging technique is a good alternative to computational temporal ghost imaging and can find numerous applications in the areas mentioned in the Introduction.

\section*{Acknowledgments}
This work was supported by QuantERA II network (project EXTRASENS) receiving funds from Horizon 2020 Framework Programme of the European Union, Grant No. 101017733. It was funded by the Bundesministerium für
Forschung, Technologie und Raumfahrt (Germany), Grant No. 13N16935, the Research Council of Finland, decision 361115, the Horizon Europe MSCA FLORIN Project 101086142, and the National Natural Science Foundation of China, Grant No.12174171.

\appendix
\section{PSF parameters in the general case \label{sec:app0}}

Introducing the shortcuts $s=\tau/T$, $q=\sigma/w$, $\varepsilon= \sqrt{1-\rho_I^2}$, $\mu=\sqrt{1-\rho_\gamma^2}$, with the help of Wolfram Mathematica 14, we obtain the following expressions for the parameters of the PSF, Eq. (\ref{generalPSF}):
\begin{align}\label{vGen}
&v=\frac{wD}{2T\left[sq \left(\varepsilon^2-2 \right)\rho_\gamma+\left( q^2+s^2+4\varepsilon^2 \right) \rho_I \right]},
\end{align}

\begin{align}\label{tauGen}
\tau_\text{PSF}=\frac{T\left[2\varepsilon^2 \left(2\varepsilon^2+q^2 \right) +s^2 \left(2\varepsilon^2+q^2\mu^2 \right) \right]^{\frac12}}{\sqrt{D}},
\end{align}
\begin{align}\label{wGen}
&w_\text{PSF}=\frac{w\sqrt{D}}{[s^2 \left( q^2\mu^2+4 \right)-8s q\rho_I\rho_\gamma+4q^2 +8\varepsilon^2 ]^{\frac12}},    
\end{align}

\begin{align}\label{PmaxGen}
&P_\text{max}=\frac{2\pi\eta_T\eta_R I_0^2\tau\sigma\varepsilon^2\mu}{\left[2\varepsilon^2 \left( 2\varepsilon^2+q^2 \right) +s^2 \left( 2\varepsilon^2 +q^2\mu^2 \right)\right]^{\frac12}},
\end{align}
where 
\begin{equation}\label{D}
D=s^2 \left( q^2\mu^2+2 \right) +8\varepsilon^2 +2q^2 \left( 1+\varepsilon^2 \right) - 4sq\rho_I\rho_\gamma
\end{equation}
is a dimensionless parameter.

\section{PSF parameters in the case of full temporal coherence \label{sec:app1}}

From Eq. (\ref{LambdaTinf}), we find 
\begin{equation}
\label{InvLambdaTinf}
\mathbf{R}^{-1} = 
\begin{bmatrix} 
 \frac{w^2\sigma^2\varepsilon^2}{\sigma^2 + 2w^2 \varepsilon^2} & 0 \\
 0 & T^2 \varepsilon^2
\end{bmatrix},
\end{equation}
and from Eq. (\ref{aTinf}),
\begin{eqnarray}\label{aLa}
\frac12\mathbf{a}^T \mathbf{R}^{-1}\mathbf{a} &=& \frac{x_2^2\rho_I^2}{2w^2\varepsilon^2} + \frac{2x_2^2 w^2 \varepsilon^2}{\sigma^2\left( \sigma^2 +2w^2\varepsilon^2 \right)}
\\ \nonumber 
&+& \frac{2x_2\tilde{t}_1w\rho_I}{T\left( \sigma^2 +2w^2\varepsilon^2 \right)} + \frac{\rho^2_I \tilde{t}_1^2 \sigma^2}{2T^2\varepsilon^2 \left( \sigma^2 +2w^2\varepsilon^2 \right)}.
\end{eqnarray}
Adding Eq. (\ref{bTinf}) to this expression, we obtain
\begin{equation}\label{QviaC}
\frac12\mathbf{a}^T \mathbf{R}^{-1}\mathbf{a} + b = -C_{xx}x_2^2-2C_{tx}x_2\tilde{t}_1-C_{tt}\tilde{t}_1^2,
\end{equation}
where
\begin{eqnarray}\label{Cxx}
C_{xx} &=& \frac{\sigma^2+2w^2+2w^2\varepsilon^2}{2w^2\left( \sigma^2 +2w^2\varepsilon^2 \right)},
\\\label{Ctt}
C_{tt} &=& \frac{\sigma^2+2w^2}{2T^2\left( \sigma^2 +2w^2\varepsilon^2 \right)},
\\\label{Ctx}
C_{tx} &=& -\frac{w\rho_I}{T\left( \sigma^2 +2w^2\varepsilon^2 \right)}.
\end{eqnarray}

We can easily verify that both eigenvalues of the coefficient matrix are positive, that is, $C_{tt}+C_{xx}>0$ and $C_{tt}C_{xx}-C_{tx}^2>0$, when $\sigma>0$ or $\varepsilon>0$.

Representing the PSF in the form of Eq. (\ref{generalPSF}), we obtain its parameters, Eqs. (\ref{v}) -- (\ref{Pmax}), from the elements of the coefficient matrix, Eqs. (\ref{Cxx}) -- (\ref{Ctx}). Note that the same parameters can be obtained from Eqs. (\ref{vGen}) -- (\ref{PmaxGen}) in the limit $s\to+\infty$, $\mu\to1$.

\bibliography{Bib-TemporalGI-2025}
\end{document}